\documentclass[preprint,showkeys,lengthcheck,twocolumn,nofootinbib,notitlepage,floatfix,superscriptaddress]{revtex4-2}
\usepackage{graphics,graphicx}
\usepackage{amsmath, amssymb}
\usepackage{multirow}
\usepackage{color}
\usepackage{verbatim}
\usepackage{hyperref}
\usepackage[normalem]{ulem}  
\usepackage{ulem}
\usepackage{color}
\usepackage{cancel}
\usepackage{mathtools}
\usepackage{epstopdf}

\renewcommand\sout{\bgroup\color{blue} \ULdepth=-.5ex \ULset}
\def\slashchar#1{\setbox0=\hbox{$#1$}  
\dimen0=\wd0     
\setbox1=\hbox{/} \dimen1=\wd1  
\ifdim\dimen0>\dimen1   
\rlap{\hbox to \dimen0{\hfil/\hfil}} 
#1     
\else     
\rlap{\hbox to \dimen1{\hfil$#1$\hfil}} 
/      
\fi}

\newcommand{\dd}{\mathrm{d}}
\newcommand{\pp}{\partial}

\begin{document}

\title{Fluctuations and correlations of baryonic chiral partners}

\date{\today}
\author{Volker Koch}
\address{Nuclear Science Division, Lawrence Berkeley National Laboratory, 1 Cyclotron Road, Berkeley, CA 94720, USA}
\address{ExtreMe Matter Institute EMMI,
GSI Helmholtzzentrum für Schwerionenforschung, Planckstraße 1,
64291 Darmstadt, Germany}
\author{Micha\l{} Marczenko}
\email{michal.marczenko@uwr.edu.pl}
\address{Incubator of Scientific Excellence - Centre for Simulations of Superdense Fluids, University of Wroc\l{}aw, plac Maksa Borna 9, PL-50204 Wroc\l{}aw, Poland}
\author{Krzysztof Redlich}
\address{Institute of Theoretical Physics, University of Wroc\l{}aw, plac Maksa Borna 9, PL-50204 Wroc\l{}aw, Poland}
\address{Polish Academy of Sciences PAN, Podwale 75, 
PL-50449 Wroc\l{}aw, Poland}
\author{Chihiro Sasaki}
\address{Institute of Theoretical Physics, University of Wroc\l{}aw, plac Maksa Borna 9, PL-50204 Wroc\l{}aw, Poland}
\address{International Institute for Sustainability with Knotted Chiral Meta Matter (WPI-SKCM$^2$), Hiroshima University, Higashi-Hiroshima, Hiroshima 739-8526, Japan}

\begin{abstract}
The exploration of critical phenomena in phase transitions of strongly interacting matter governed by Quantum Chromodynamics (QCD) is one of the goals of present ultrarelativistic heavy-ion collision experiments at BNL and CERN. The key research direction is to locate the putative critical point on the phase diagram of QCD linked to the chiral symmetry restoration at finite temperature and/or density. One of the main theoretical tools used for this purpose are the fluctuations of conserved charges, such as the net-baryon number. However, due to experimental limitations, analyses of heavy-ion collision data suffer from a very doubtful basing of the net-proton number being a proxy for the total net-baryon number fluctuations. In this work, we use the parity doublet model to investigate the fluctuations of the net-baryon number density in hot and dense hadronic matter. The model accounts for chiral criticality within the mean-field approximation. We focus on the qualitative properties and systematics of the first- and second-order susceptibility of the net-baryon number density, and their ratios for nucleons of positive and negative parity, as well as their correlator. We show that the fluctuations of the positive-parity nucleon do not necessarily reflect the fluctuations of the total net-baryon number density at the phase boundary of the chiral phase transition. We also investigate the non-trivial structure of the correlator. Furthermore, we discuss and quantify the differences between the fluctuations of the net-baryon number density in the vicinity of the chiral and liquid-gas phase transition in nuclear matter. We indicate a possible relevance of our results with the interpretation of the experimental data on net-proton number fluctuations in heavy-ion collisions.
\end{abstract}
\maketitle

\section{Introduction}
\label{sec:intro}

One of the prominent tasks within high-energy physics is to unveil the phase diagram of Quantum Chromodynamics (QCD), the theory of strong interactions. Due to great activity in the field, significant progress has been made from both the theoretical and experimental sides. From ab initio lattice QCD (LQCD) calculations, it is now known that, at vanishing baryon density, strongly interacting matter undergoes a smooth chiral symmetry restoration transition from hadronic matter to quark-gluon plasma (QGP) at $T_c\approx155~$MeV~\cite{Bazavov:2014pvz, Borsanyi:2018grb, Bazavov:2017dus, Bazavov:2020bjn, Bazavov:2020bjn, Aoki:2006we}. However, the applicability of the LQCD methods at high baryon densities ceases, due to a well-known sign problem. Effective models, such as the linear sigma~\cite{Bowman:2008kc, Ferroni:2010ct} or Nambu--Jona-Lasinio (NJL)~\cite{Klevansky:1992qe, Buballa:2003qv} models predict a first-order phase transition at low temperature. Its existence would imply the presence of a putative critical endpoint (CP) on the QCD phase diagram. Throughout recent years experimental attempts were made to locate it on the phase diagram of QCD. Despite enormous experimental effort within the beam energy scan (BES) programs at the Relativistic Heavy Ion Collider (RHIC) at BNL~\cite{STAR2010} and the Super Proton Synchrotron (SPS) at CERN~\cite{Mackowiak-Pawlowska:2020glz}, this pressing issue remains unresolved (for a recent review see~\cite{Bzdak:2019pkr}).

One of the tools used in the experimental searches of the critical point are fluctuations and correlations of conserved charges. They are known to be propitious theoretical observables in search of critical behavior at the QCD phase boundary~\cite{Stephanov:1999zu, Asakawa:2000wh, Hatta:2003wn, Friman:2011pf} and chemical freeze-out in the heavy-ion collisions (HIC)~\cite{Bazavov:2012vg, Borsanyi:2014ewa, Karsch:2010ck, Braun-Munzinger:2014lba, Vovchenko:2020tsr, Braun-Munzinger:2020jbk}. In particular, fluctuations of conserved charges have been proposed to probe the QCD critical point, as well as the remnants of the $O(4)$ criticality at vanishing and finite net-baryon densities~\cite{Friman:2011pf, Stephanov:2011pb, Karsch:2019mbv, Braun-Munzinger:2020jbk, Braun-Munzinger:2016yjz}. 

Non-monotonic behavior is also expected for various ratios of the cumulants of the net-baryon number. Recently, results from BES-I, which covered $\sqrt{s_{\rm NN}}=7.7-200~$GeV, have shown indications of a non-monotonic behavior of the forth-to-second cumulant ratio of the net-proton multiplicity distributions in central Au+Au collisions~\cite{STAR:2020tga}. However, more data and higher statistics at low collision energies are needed to draw firm conclusions.

However, due to experimental limitations, only charged particles created in a heavy-ion collision have a fair chance of being measured by the detector. The net-proton number fluctuations can be faithfully assumed to be roughly half of the net-nucleon number fluctuations. This is a fair assumption since isospin correlations are expected to be small~\cite{Fukushima:2014lfa}, due to, e.g., isospin randomization. However, an assumption that the fluctuations of the net-proton (or net-nucleon) number should reflect the overall fluctuations of the net-baryon number is very doubtful. The relation and differences between net-baryon and net-proton number fluctuations have not yet been explored in theoretical models that consider dynamical chiral symmetry restorations in a strongly interacting medium.

One of the consequences of the restoration of chiral symmetry is the emergence of parity doubling around the chiral crossover. This has been recently observed in LQCD calculations in the spectrum of low-lying baryons around the chiral crossover~\citep{Aarts:2015mma, Aarts:2017rrl, Aarts:2018glk}. The masses of the positive-parity baryonic ground states are found to be rather weakly temperature-dependent, while the masses of negative-parity states drop substantially when approaching the chiral crossover temperature. The parity doublet states become almost degenerate with a finite mass in the vicinity of the chiral crossover. Such properties of the chiral partners can be described in the framework of the parity doublet model~\citep{Detar:1988kn, Jido:1999hd, Jido:2001nt}. The model has been applied to the vacuum phenomenology of QCD, hot and dense hadronic matter, as well as neutron stars~\citep{Dexheimer:2007tn, Gallas:2009qp, Paeng:2011hy, Sasaki:2011ff, Gallas:2011qp, Zschiesche:2006zj, Benic:2015pia, Marczenko:2017huu, Marczenko:2018jui, Marczenko:2019trv, Marczenko:2020wlc, Marczenko:2020jma, Marczenko:2021uaj, Marczenko:2022hyt, Mukherjee:2017jzi, Mukherjee:2016nhb, Dexheimer:2012eu, Steinheimer:2011ea, Weyrich:2015hha, Sasaki:2010bp, Yamazaki:2018stk, Yamazaki:2019tuo, Ishikawa:2018yey, Steinheimer:2010ib, Giacosa:2011qd, Motohiro:2015taa, Minamikawa:2020jfj, Kong:2023nue}.

In this paper, we apply the parity doublet model to calculate the cumulants and susceptibilities of the net-baryon number distribution. Specifically, we focus on the fluctuations of individual parity channels and correlations among them. Their qualitative behavior is examined near the chiral, as well as the nuclear liquid-gas phase transitions.

The differences in the qualitative critical behavior of opposite parity states were shown to be non-trivial, e.g., the difference of the sign of contributing terms to the overall fluctuations that are linked to the positive- and negative-parity states~\cite{Marczenko:2023ohi}. The decomposition performed in this study, however, cannot be interpreted in terms of cumulants of the baryon number. In this work, we extend this analysis by explicitly evaluating the fluctuations in the individual parity channels, as well as the correlation among them.  

This work is organized as follows. In Sec.~\ref{sec:pd_model}, we introduce the hadronic parity doublet model. In Sec.~\ref{sec:fluct}, we introduce the cumulants and susceptibilities of the net-baryon number. In Sec.~\ref{sec:results}, we present our results. Finally, Sec.~\ref{sec:conclusions} is devoted to the summary of our findings.

\section{Parity doublet model}
\label{sec:pd_model}

The hadronic parity doublet model for the chiral symmetry restoration~\cite{Detar:1988kn, Jido:2001nt, Jido:1999hd} is composed of the baryonic parity doublet and mesons as in the Walecka model~\cite{Walecka:1974qa}. The spontaneous chiral symmetry breaking yields the mass splitting between the two fermionic parity partners. In this work, we consider a system with $N_f = 2$; hence, relevant for this study are the positive-parity nucleons and their negative-parity partners. The fermionic degrees of freedom are coupled to the chiral fields ($\sigma$, $\boldsymbol\pi$) and the isosinglet vector field ($\omega_\mu$).

To investigate the properties of strongly interacting matter, we adopt a mean-field approximation. Rotational invariance requires that the spatial component of the $\omega_\mu$ field vanishes, namely, $\langle \boldsymbol \omega \rangle = 0$\footnote{Since $\omega_0$ is the only non-zero component in the mean-field
approximation, we simply denote it by $\omega_0 \equiv \omega$.}. Parity conservation on the other hand dictates $\langle\boldsymbol\pi\rangle=0$. The mean-field
thermodynamic potential of the parity doublet model reads~\cite{Marczenko:2023ohi}\footnote{Assuming isospin symmetric system.}\footnote{We present the details of the Lagrange formulation of the model in Appendix~\ref{sec:PDM_appendix}.}
\begin{equation}\label{eq:thermo_potential}
\Omega = \Omega_+ + \Omega_- + V_\sigma + V_\omega \textrm,
\end{equation}
with
\begin{equation}\label{eq:kinetic_thermo}
\Omega_\pm = \gamma_\pm \int\frac{\dd^3 p}{(2\pi)^3}\; T \left[ \ln\left(1 - f_\pm\right) + \ln\left(1 - \bar f_\pm\right) \right]\textrm,
\end{equation}
where $\gamma_\pm = 2\times 2$ denotes the spin-isospin degeneracy factor for both parity partners, and $f_\pm$  $(\bar f_\pm)$ is the particle (antiparticle) Fermi-Dirac distribution function,
\begin{equation}\label{eq:fermi_dist_nucleon}
\begin{split}
f_\pm = \frac{1}{1+ e^{\left(E_\pm - \mu_N\right)/T}} \textrm,\\
\bar f_\pm = \frac{1}{1+ e^{\left(E_\pm + \mu_N\right)/T}}\textrm, \\
\end{split}
\end{equation}
where $T$ is the temperature, the dispersion relation $E_\pm = \sqrt{\boldsymbol p^2 + m_\pm^2}$,  and the effective baryon chemical potential $\mu_N = \mu_B - g_\omega \omega$. The mean-field potentials read
\begin{subequations}\label{eq:potentials_parity_doublet}
\begin{align}
    V_\sigma &= -\frac{\lambda_2}{2}\Sigma + \frac{\lambda_4}{4}\Sigma^2 - \frac{\lambda_6}{6}\Sigma^3- \epsilon\sigma \textrm,\label{eq:potentials_sigma}\\
    V_\omega &= -\frac{m_\omega^2 }{2}\omega^2\textrm.
\end{align}
\end{subequations}
where $\Sigma = \sigma^2 + \boldsymbol\pi^2$, $\lambda_2 = \lambda_4f_\pi^2 - \lambda_6f_\pi^4 - m_\pi^2$, and $\epsilon = m_\pi^2 f_\pi$. $m_\pi$ and $m_\omega$ are the $\pi$ and $\omega$ meson masses, respectively, and $f_\pi$ is the pion decay constant.

The masses of the positive- and negative-parity baryonic chiral partners, $N_\pm$, are given by 
\begin{equation}\label{eq:doublet_masses}
    m_\pm = \frac{1}{2}\left(\sqrt{a^2\sigma^2 + 4m_0^2} \mp b\sigma\right) \textrm,
\end{equation}
where $a$, $b$ are combinations of Yukawa coupling constants~\cite{Marczenko:2023ohi}, and $m_0$ is the chirally invariant mass parameter. We note that in the parity doublet model, the chiral symmetry breaking yields the mass splitting between the chiral partners. Therefore, the order parameter for the chiral symmetry breaking is the mass difference, $m_- - m_+ = b\sigma$.

\begin{table*}[t!]\begin{center}\begin{tabular}{|c|c|c|c|c|c|c|c|c|c|c|}
\hline
$m_0~$[GeV] & $m_+~$[GeV] & $m_-~$[GeV] & $m_\pi~$[GeV] & $f_\pi~$[GeV] & $m_\omega~$[GeV] & $\lambda_4$ & $\lambda_6f_\pi^2$ & $g_\omega$ & $a$ & $b$ \\ \hline\hline
0.750 & 0.939   & 1.500  & 0.140     & 0.93      & 0.783        &   28.43    &   11.10       &    6.45    & 20.68 & 6.03 \\ \hline
\end{tabular}\end{center}
\caption{Physical inputs in matter-free space and the model parameters used in this work. See Sec.~\ref{sec:pd_model} for details.}
\label{tab:vacuum_params}
\end{table*}

In-medium profiles of the mean fields are obtained by extremizing the thermodynamic potential, Eq.~\eqref{eq:thermo_potential}, leading to the following gap equations:
\begin{equation}
\begin{split}\label{eq:gap_eqs}
0=\frac{\partial \Omega}{\partial \sigma} &= \frac{\partial V_\sigma}{\partial \sigma} + s_+ \frac{\partial m_+}{\partial \sigma} + s_- \frac{\partial m_-}{\partial \sigma} \textrm,\\
0=\frac{\partial \Omega}{\partial \omega} &= \frac{\partial V_\omega}{\partial \omega} + g_\omega \left(n_+ + n_-\right) \textrm,
\end{split}
\end{equation}
where the scalar and vector densities are
\begin{equation}
s_\pm = \gamma_\pm \int \frac{\mathrm{d}^3 p}{\left(2\pi\right)^3} \frac{m_\pm}{E_\pm}\left(f_\pm + \bar f_\pm\right)
\end{equation}
and
\begin{equation}
n_\pm = \gamma_\pm \int \frac{\mathrm{d}^3 p}{\left(2\pi\right)^3}\left(f_\pm - \bar f_\pm\right) \textrm,
\end{equation}
respectively.

In the grand canonical ensemble, the net-baryon number density can be calculated as follows:
\begin{equation}\label{eq:nb}
n_B = -\frac{\dd \Omega}{\dd \mu_B}\Bigg|_T = n_+ + n_-\textrm,
\end{equation}
where $n_\pm$ are the vector densities of the baryonic chiral partners.

The positive-parity state, $N_+$, corresponds to the nucleon $N(938)$. Its negative parity partner, $N_-$, is identified with $N(1535)$~\cite{ParticleDataGroup:2022pth}. Their vacuum masses are shown in Table~\ref{tab:vacuum_params}. The value of the parameter $m_0$ has to be chosen so that a chiral crossover is realized at finite temperature and vanishing chemical potential. The model predicts the chiral symmetry restoration to be a crossover for $m_0\gtrsim 700~$MeV. Following the previous studies of the \mbox{parity-doublet-based} models~\cite{Dexheimer:2007tn,Gallas:2009qp,Paeng:2011hy, Sasaki:2011ff, Gallas:2011qp, Zschiesche:2006zj, Benic:2015pia, Marczenko:2017huu, Marczenko:2018jui, Marczenko:2019trv, Marczenko:2020wlc, Marczenko:2020jma, Marczenko:2023ohi, Mukherjee:2017jzi, Mukherjee:2016nhb, Dexheimer:2012eu, Steinheimer:2011ea, Weyrich:2015hha, Sasaki:2010bp, Yamazaki:2018stk,Yamazaki:2019tuo, Ishikawa:2018yey, Steinheimer:2010ib, Giacosa:2011qd,Motohiro:2015taa,Minamikawa:2020jfj}, as well as recent lattice QCD results~\cite{Aarts:2017rrl, Aarts:2018glk,Aarts:2015mma}, we choose a rather large value, $m_0=750$~MeV. We note, however, that the results presented in this work qualitatively do not depend on the choice of $m_0$, as long as the chiral crossover appears at $\mu_B=0$. The parameters $a$ and $b$ are determined by the aforementioned vacuum nucleon masses and the chirally invariant mass $m_0$ via Eq.~\eqref{eq:doublet_masses}. The remaining parameters: $g_\omega$, $\lambda_4$ and $\lambda_6$, are fixed by the properties of the nuclear ground state at zero temperature, i.e., the saturation density, binding energy, and compressibility parameter at $\mu_B=923~$MeV. The constraints are as follows:
\begin{subequations}
\begin{align}
n_B &= 0.16~\textrm{fm}^{-3}\textrm,\\
E/A - m_+ &= -16~\textrm{MeV}\textrm,\\
K = 9n^2_B\frac{\pp^2\left( E/A \right)}{\pp n_B^2} &= 240~\textrm{MeV}\textrm.\label{eq:compres}
\end{align}
\end{subequations}
We note that the six-point scalar interaction term in Eq.~\eqref{eq:potentials_sigma} is essential to reproduce the empirical value of the compressibility in Eq.~\eqref{eq:compres}~\citep{Motohiro:2015taa}.

\begin{figure}
    \centering
    \includegraphics[width=0.95\linewidth]{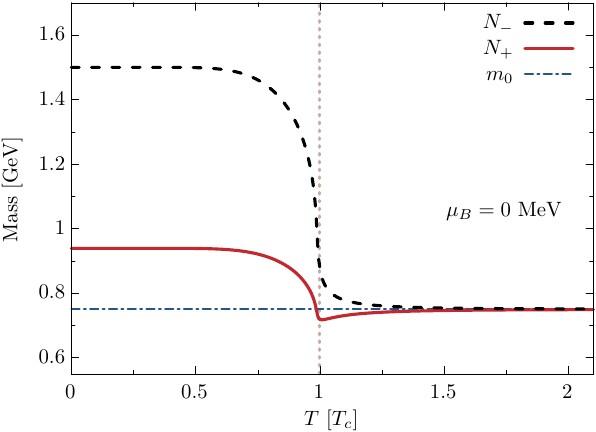}
    \caption{Masses of the baryonic chiral partners at finite temperature and vanishing baryon chemical potential. The temperature is normalized to the chiral crossover temperature, $T_c$, at $\mu_B=0$. The dotted, blue line shows the chirally invariant mass, $m_0$. The vertical line marks the chiral crossover transition.}
    \label{fig:m_T}
\end{figure}

The compilation of the parameters used in this paper is found in Table~\ref{tab:vacuum_params}. For this set of parameters, we obtain the pseudo-critical temperature of the chiral crossover at vanishing baryon chemical potential, $T_c=209~$MeV. In Fig.~\ref{fig:m_T} we show the temperature dependence of the masses of the chiral partners. At low temperatures,  chiral symmetry is broken and they have different masses. As chiral symmetry gets restored, their masses converge towards the chirally invariant mass $m_0$. The mass of the $N_-$ monotonically decreases towards $m_0$. On the other hand, the mass of $N_+$ develops a shallow minimum close to the chiral restoration and converges to $m_0$ from below. The derivatives of $m_\pm$ can be readily calculated from Eq.~\eqref{eq:doublet_masses}, namely
\begin{equation}\label{eq:m_ds}
    \frac{\pp m_\pm}{\pp\sigma} = \frac{1}{2}\left(\frac{a^2\sigma}{\sqrt{a^2\sigma^2 + 4m_0^2}}\mp b \right) \textrm.
\end{equation}
Note that for the positive-parity state, a minimum value of the mass, $m_+^{\rm min}$, exists at
\begin{equation}
\sigma_{\rm min} = \frac{2b m_0}{a \sqrt{a^2-b^2}}\textrm,
\end{equation}
while the mass of the negative-parity state monotonically decreases with $\sigma$ as the chiral symmetry gets restored. We also note that $\sigma_{\rm min} > 0$; Thus, the positive-parity state attains a minimum mass for any choice of $m_0>0$~\cite{Marczenko:2023ohi}. 

\begin{figure}
    \centering
    \includegraphics[width=0.95\linewidth]{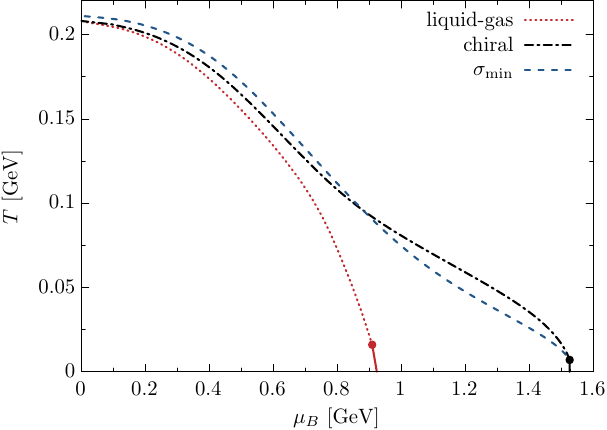}
    \caption{Phase diagram obtained in the parity doublet model. Shown are the liquid-gas (red, solid/dotted line) and chiral (black, solid/dashed-dotted line) phase transition/crossover lines. Circles indicate critical points below which the transitions are of the first order. The lines are obtained from the minima of $\pp \sigma / \pp \mu_\pm$ (see text for details). The blue, dashed line shows the line where the mass of the positive-parity state has a minimum (see text for details).}
    \label{fig:phase_diag}
\end{figure}

At low temperatures, the model predicts sequential first-order nuclear liquid-gas and chiral phase transitions with critical points located at $T_{\rm lg} = 16~$MeV, $\mu_B=909~$MeV, ($n_B=0.053~\textrm {fm}^{-3}=0.33n_0$) and $T_{\rm ch} = 7~$MeV, $\mu_B=1526~$MeV ($n_B=1.25~\textrm{fm}^{-3}=7.82n_0$), respectively. In Fig.~\ref{fig:phase_diag}, we show the parity doublet model phase diagram. At low temperature, the nuclear liquid-gas and chiral phase transitions are sequential. As temperature increases, they combine and form a single crossover transition at vanishing baryon chemical potential. We note that the exact location of the chiral phase transition at low temperature depends on, e.g., the mass of the negative-parity state~\cite{Zschiesche:2006zj}. At zero temperature it is expected that it occurs roughly at $\mu_B \sim m_-$. 

We note that the minimum of $m_+$ is obtained for any trajectory from chirally broken to chirally symmetric phase. Remarkably,  $\sigma_{\rm min}$ is reached at $T$ and $\mu_B$ which are close to the chiral phase boundary (see Fig.~\ref{fig:phase_diag}). We emphasize that the properties discussed in this work are expected to appear independently of the position of the chiral critical point on the phase diagram. Although the dependence of $m_+$ on $\sigma$ is not universal and model dependent, we stress that the calculations with the functional renormalization group techniques preserve the same in-medium behavior~\cite{Tripolt:2021jtp}. At present, only the first-principle LQCD calculations can provide a reliable answer.

We also note that the parity doublet model, considered here,  is only valid in describing chiral symmetry restoration from the side of the hadronic phase. Thus, it does not contain any information about the transition to deconfined quark matter. However, at vanishing and small chemical potential the smooth deconfinement transition is simultaneous with the restoration of chiral symmetry ~\cite{Bazavov:2014pvz, Borsanyi:2018grb, Bazavov:2017dus, Bazavov:2020bjn, Bazavov:2020bjn, Aoki:2006we}. Therefore, the applicability of the parity doublet model should be adequate up to the chiral crossover temperature and density, although the interplay between chiral symmetry restoration and deconfinement in high-density matter is still not known.

In the next section, we discuss the general structure of the second-order susceptibilities of the net-baryon number density for positive- and negative-parity chiral partners to quantify their roles near the second-order phase transition at finite density.

\section{Cumulants and susceptibilities of the net-baryon number}\label{sec:fluct}

For a system consisting of $N_B = N_+ + N_-$ baryons with $N_\pm$ being the net number of positive/negative-parity baryons, the mean can be calculated as
\begin{equation}
   \langle N_B \rangle \equiv \kappa_1^B = \kappa_1^+ + \kappa_1^- \rm,
\end{equation}
and the variance,
\begin{equation}
    \langle \delta N_B \delta N_B\rangle \equiv \kappa_2^B  = \kappa_2^{++} + \kappa_2^{--} + 2\kappa_2^{+-} \rm,
\end{equation}
where
\begin{equation}
\begin{split}
    \kappa_1^\alpha &= \langle N_\alpha \rangle \rm, \\
    \kappa_2^{\alpha\beta} &= \langle\delta N_\alpha \delta N_\beta\rangle \rm.
\end{split}
\end{equation}
Notably $\kappa_1^\pm$, $\kappa_2^{++}$ and $\kappa_2^{--}$ are the cumulants of the $N_+$ and $N_-$ distributions; $\kappa_2^{+-}$ is the correlation between $N_+$ and $N_-$.

In general, the cumulants of the baryon number are defined as
\begin{equation}
    \kappa_n^B \equiv T^n \frac{\dd^n \log{\mathcal Z}}{\dd \mu_B^n}\Bigg|_T \rm,
\end{equation}
where $\mathcal Z$ is the partition function. Because the thermodynamic potential $\Omega$ is related to the grand-canonical partition function through $\Omega = -T\log{\mathcal Z} / V$, one may relate the cumulants with the susceptibilities of the net-baryon number in the following way
\begin{equation}\label{eq:def_x2_kn}
    \kappa_n^B = V T^3\chi_n^B \rm,
\end{equation}
where $V$ is the volume of the system and 
\begin{equation}\label{eq:chi_def}
    \chi_n^B \equiv -\frac{\dd^n \hat\Omega}{\dd \hat\mu_B^n}\Bigg|_T \rm,
\end{equation}
with $\hat \Omega = \Omega/T^4$ and $\hat \mu_B = \mu_B/T$. For example, $\kappa_1^B = V \chi_1^B = V n_B = \langle N_B\rangle$ is the mean of the baryon number. We note that $\langle N_B\rangle = \langle N_+\rangle + \langle N_-\rangle$ is the sum of the means of the net number of particles with a given parity; thus $\kappa_1^B = \kappa_1^+ + \kappa_1^-$, where $\kappa_1^\alpha = \langle N_\alpha \rangle$.

To be able to connect the individual cumulants $\kappa_n^{\alpha\beta}$ to susceptibilities, we need to rewrite the mean-field thermodynamic potential in terms of newly defined chemical potentials, $\mu_\pm$ for positive- and negative-parity states as follows:
\begin{equation}\label{eq:thermo_pm}
\begin{split}
    \Omega &=\Omega_+\left(\mu_+, T, \sigma\left(\mu_+, \mu_-\right), \omega\left(\mu_+, \mu_-\right)\right)\\
    &+ \Omega_-\left(\mu_-, T, \sigma\left(\mu_+, \mu_-\right), \omega\left(\mu_+, \mu_-\right)\right) \\
    &+ V_\sigma(\sigma\left(\mu_+, \mu_-\right)) + V_\omega(\omega\left(\mu_+, \mu_-\right))\rm.
\end{split}
\end{equation}
Such a separation into separate chemical potentials is possible in the mean field approximation which is a single particle theory (see detailed discussion in~\cite{Garcia:2018iib}). To be thermodynamically consistent, one needs to set $\mu_\pm = \mu_N = \mu_B - g_\omega \omega$ at the end of the calculations and before numerical evaluation. We note that $\mu_\pm$ are independent variables. The net-baryon density is then given as
\begin{equation}
    n_B = n_+ + n_-\rm,
\end{equation}
where $n_\pm$ are the net densities given by
\begin{equation}\label{eq:dens_pm}
\begin{split}
n_\pm &= -\frac{\dd \Omega}{\dd \mu_\pm}\Bigg|_{T,\mu_\pm=\mu_N}  =\\
&- \frac{\pp \Omega}{\pp \mu_\pm} - \frac{\pp \Omega}{\pp \sigma}\frac{\pp \sigma}{\pp \mu_\pm} - \frac{\pp \Omega}{\pp \omega}\frac{\pp \omega}{\pp \mu_\pm}= - \frac{\pp \Omega}{\pp \mu_\pm}\rm,
\end{split}
\end{equation}
The last equality holds due to the stationary conditions. We stress that the derivative should be taken not only at constant temperature but also at $\mu_+=\mu_-=\mu_N$.

Given that $\mu_\pm$ are independent, one recognizes that Eq.~\eqref{eq:dens_pm} agrees with the definition in Eq.~\eqref{eq:nb}. Likewise, the second-order susceptibility can be expressed as follows
\begin{equation}\label{eq:x2_sum}
\chi_2^B = \chi_2^{++} + \chi_2^{--} + 2\chi_2^{+-} \rm,
\end{equation}
where $\chi_{++}~(\chi_{--})$ are the susceptibilities of the positive-(negative-) parity and $\chi_{+-}$ gives the correlations between them, i.e., correlations between vector densities. The individual terms in the above equation are given as follows
\begin{equation}\label{eq:chi_2_ab}
    \chi_2^{\alpha\beta} =  \frac{1}{VT^3} \kappa_2^{\alpha\beta} = -\frac{\dd^2 \hat\Omega}{\dd \hat\mu_\alpha \dd\hat\mu_\beta}\Bigg|_{T,\mu_\alpha=\mu_\beta=\mu_N} \rm,
\end{equation}
where $\hat\mu_{x} = \mu_{x}/T$, and $x = \alpha, \beta$ correspond to the particle species and $\mu_\alpha$'s correspond to their effective chemical potentials $\mu_\pm$.
Detailed derivation of the cumulants $\kappa_2^{\alpha\beta}$ is presented in Appendix~\ref{sec:appendix_0}. We notice that, under the mean-field approximation, $\chi_2^{\alpha\beta} = \chi_2^{\beta\alpha}$,  thus $\chi_2^{+-} = \chi_2^{-+}$. Furthermore, we assume isospin symmetry, thus $\chi_2^{++}$ is the net-nucleon number susceptibility. Consequently, the susceptibility of the net-proton number density is $\chi_2^{pp} \approx 1/2 \chi_2^{++}$. This is a fair assumption since isospin correlations are expected to be small~\cite{Fukushima:2014lfa}.

\begin{figure}
    \centering 
    \includegraphics[width=0.95\linewidth]{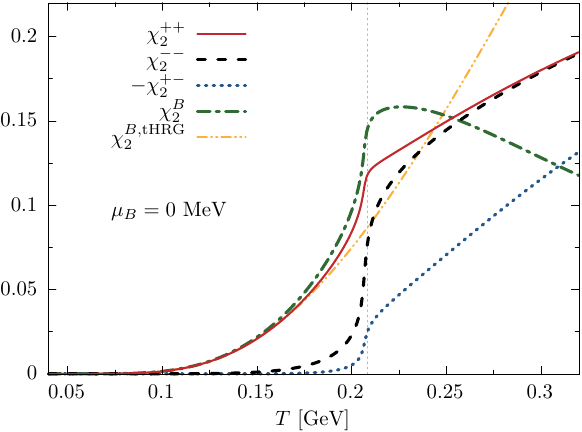}
    \caption{Susceptibilities, $\chi_2^{\alpha\beta}$, at vanishing baryon chemical potential. Shown are also the net-baryon number susceptibility $\chi_2^B$ and the corresponding result, $\chi_2^{B, \rm tHRG}$ obtained in the tHRG model. We note that the correlator, $\chi_2^{+-}$, is shown with the negative sign. The vertical, dotted line marks the chiral phase transition.}
    \label{fig:x2_T}
\end{figure}

\begin{figure*}
    \centering
    \includegraphics[width=.475\linewidth]{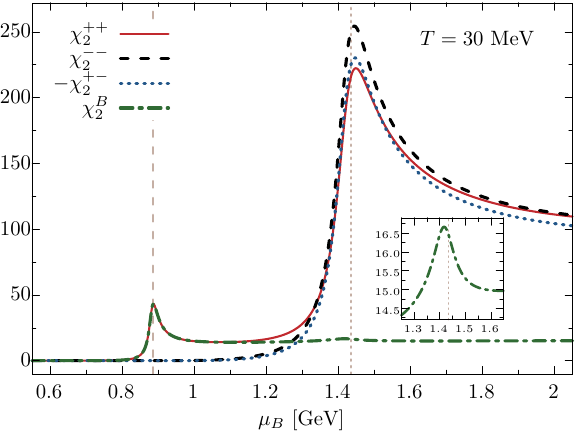}\;\;\;\;
    \includegraphics[width=.475\linewidth]{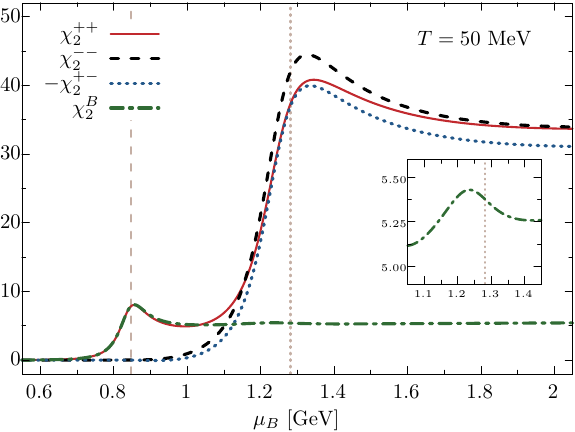}
    \includegraphics[width=.475\linewidth]{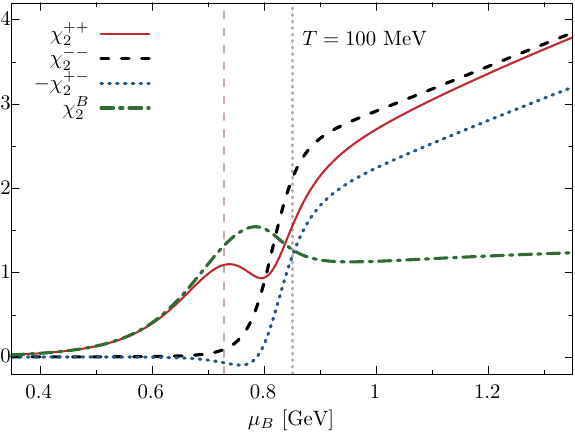}\;\;\;
    \includegraphics[width=.475\linewidth]{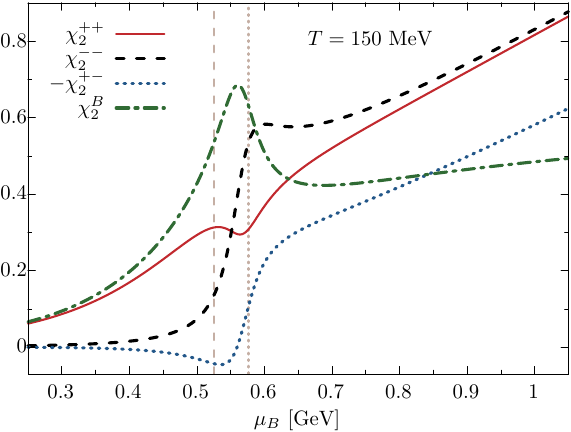}
    \caption{Susceptibilities, $\chi_2^{\alpha\beta}$, at different temperatures. Also shown is the net-baryon number susceptibility, $\chi_2^B$. We note that the correlator, $\chi_2^{+-}$, is shown with the negative sign. The dashed and dotted vertical lines mark baryon chemical potentials for the liquid-gas and chiral crossover transitions, respectively. The inset figures in the top panel show $\chi_2^B$ in the vicinity of the chiral crossover transition.}
    \label{fig:x2_fixed_t_panel}
\end{figure*}

Event-by-event cumulants and correlations are extensive quantities. They depend on the volume of the system and its fluctuations, which are unknown in heavy-ion collisions. The volume dependence, however, can be cancelled out by taking the ratio of cumulants. Therefore, it is useful to define ratios of the cumulants of the baryon number, which may also be expressed through susceptibilities,
\begin{equation}
    R_{n,m}^B \equiv \frac{\kappa_n^B}{\kappa_m^B} = \frac{\chi_n^B}{\chi_m^B} \rm.
\end{equation}
In the following, we focus on the ratios of the second and first-order cumulants of different parity distributions. Therefore, it is useful to define
\begin{equation}
    R_{2,1}^{\alpha\beta} \equiv \frac{\kappa_2^{\alpha\beta}}{\sqrt{\kappa_1^\alpha \kappa_1^{\beta}}} = \frac{\chi_2^{\alpha\beta}}{\sqrt{\chi_1^\alpha \chi_1^\beta}} \rm.
\end{equation}
We note that in general the ratios, $R_{n,m}^{\alpha\beta}$, are not additive, e.g., $R_{2,1}^{++} + R_{2,1}^{--} + 2R_{2,1}^{+-} \neq R_{2,1}^B$.

\begin{figure*}
    \centering
    \includegraphics[width=.475\linewidth]{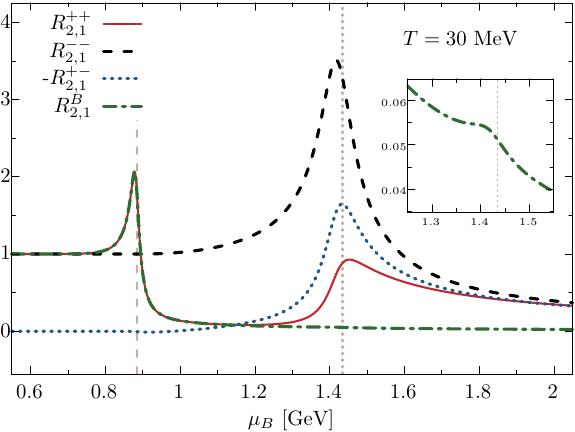}\;\;\;\;
    \includegraphics[width=.475\linewidth]{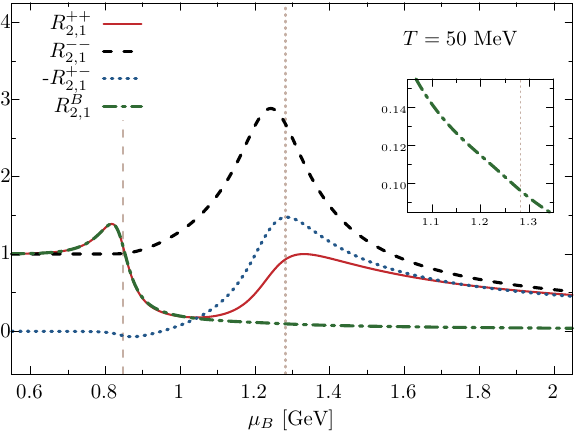}
    \includegraphics[width=.475\linewidth]{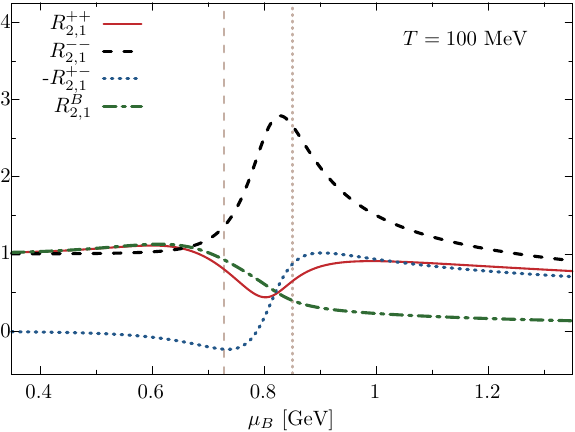}\;\;\;\;
    \includegraphics[width=.475\linewidth]{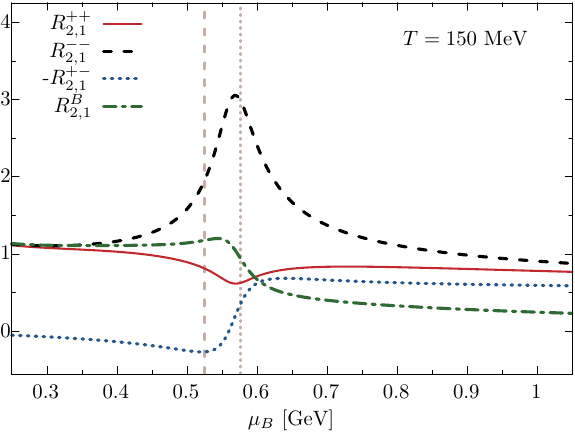}
    \caption{Scaled variances, $R_{2,1}^{\alpha\beta}$ for different temperatures. Also shown is the ratio $R_{2,1}^B$, for the net-baryon number susceptibility. We note that the ratio, $R_{2,1}^{+-}$, is shown with the negative sign. The dashed and dotted vertical lines mark baryon chemical potentials for the liquid-gas and chiral crossover transitions, respectively. In the top panel, the inset figures show $R_{2,1}^B$ in the vicinity of the chiral crossover transition.}
    \label{fig:R21_fixed_t_panel}
\end{figure*}

In the following, we will also compare our results with the truncated hadron resonance gas (tHRG) model, which in the relevant baryonic sector contains contributions only from nucleons $N_+$ and their parity partners $N_-$. The HRG model is widely used for the description of matter under extreme conditions, e.g., in the context of heavy-ion collision phenomenology~\cite{Karsch:2003vd, Karsch:2003zq, Karsch:2013naa, Andronic:2012ut, Albright:2014gva, Albright:2015uua}. Commonly used implementations of the HRG employ vacuum hadron masses in the hadronic phase and hence do not include possible in-medium effects. Several extensions of the HRG model have been proposed to quantify the LQCD EoS and various fluctuation observables. They account for consistent implementation of hadronic interactions within the S-matrix approach~\cite{Andronic:2017pug}, a more complete implementation of a continuously growing exponential mass spectrum and/or possible repulsive interactions among constituents~\cite{Majumder:2010ik, Andronic:2012ut, Albright:2014gva, Lo:2015cca, ManLo:2016pgd, Andronic:2020iyg, Vovchenko:2016rkn}. Nevertheless, it is challenging to identify the role of different in-medium effects and hadronic interactions on the properties of higher-order fluctuations of conserved charges. 

The thermodynamic potential of the tHRG model is a mixture of uncorrelated ideal gases of stable $N_\pm$ particles:
\begin{equation}\label{eq:hrg_thermo}
    \Omega^{\rm tHRG} = \sum_{x=\pm} \Omega_x \textrm,
\end{equation}
with $\Omega_x$ given by Eq.~\eqref{eq:kinetic_thermo}. The masses of $N_\pm$ are taken to be the vacuum masses (see Table~\ref{tab:vacuum_params}) and $\mu_N = \mu_B$. The net-baryon density and its susceptibility are obtained through Eqs.~\eqref{eq:nb}~and~\eqref{eq:chi_def}, respectively. Thus, in the tHRG model one has,
\begin{equation}
\chi_2^{B, \rm tHRG} = \chi_2^{++} + \chi_2^{--}\rm.    
\end{equation}

The susceptibilities introduced in Eq.~\eqref{eq:chi_2_ab}, can be evaluated analytically by differentiating Eq.~\eqref{eq:thermo_pm}. Explicit calculations yield
\begin{equation}\label{eq:x2_ab2}
\begin{split}
    \chi_2^{\alpha\beta} = &-\frac{\pp \sigma}{\pp \hat\mu_\beta} \left( \frac{\pp^2 \hat\Omega}{\pp \sigma^2}\frac{\pp \sigma}{\pp \hat\mu_\alpha} + \frac{\pp^2\hat\Omega}{\pp\sigma\pp\omega}\frac{\pp\omega}{\pp\hat\mu_\alpha} - \frac{\pp \hat n_\alpha}{\pp\sigma} \right)\\
    &-\frac{\pp\omega}{\pp\hat\mu_\beta} \left( \frac{\pp^2\hat\Omega}{\pp\omega^2}\frac{\pp \omega}{\pp\hat\mu_\alpha} + \frac{\pp^2\Omega}{\pp\sigma\pp\omega}\frac{\pp\sigma}{\pp\hat\mu_\alpha} - \frac{\pp \hat n_\alpha}{\pp\omega}\right)\\
    &+\frac{\pp\sigma}{\pp\hat \mu_\alpha}\frac{\pp \hat n_\beta}{\pp\sigma}+ \frac{\pp\omega}{\pp\hat \mu_\alpha}\frac{\pp\hat n_\beta}{\pp\omega} + \frac{\pp \hat n_\alpha}{\pp\hat \mu_\beta} \rm,
    \end{split}
\end{equation}
where $\hat n_{\alpha/\beta} = n_{\alpha/\beta} /T^3$, and $n_{\alpha/\beta}$ are the net densities defined in Eq.~\eqref{eq:dens_pm}. We note that the last term, $\pp \hat n_\alpha / \pp\hat\mu_\beta = 0$ for $\alpha \neq \beta$.

To evaluate Eq.~\eqref{eq:x2_ab2}, one needs to extract the derivatives of the mean fields w.r.t chemical potentials $\mu_\pm$. They can be carried out by differentiating the gap equations, namely
\begin{equation}
\begin{split}
    &\frac{\dd}{\dd \hat\mu_\alpha}\left(\frac{\pp \hat\Omega}{\pp \sigma}\right)\Bigg|_{T,\hat\mu_\alpha=\hat\mu_N} = 0\rm,\\
    &\frac{\dd}{\dd \hat\mu_\alpha}\left(\frac{\pp \hat\Omega}{\pp \omega}\right)\Bigg|_{T,\hat\mu_\alpha=\hat\mu_N} = 0 \rm.
\end{split}
\end{equation}
Writing them explicitly and isolating $\pp \sigma / \hat\mu_\alpha$, $\pp \omega / \pp \hat\mu_\alpha$, yields
\begin{equation}
\begin{split}
    \frac{\pp \sigma}{\pp \hat \mu_\alpha} = &\left( \frac{\frac{\pp^2 \hat \Omega}{\pp\sigma\pp\omega}}{\frac{\pp^2\hat\Omega}{\pp\omega^2}}\frac{\pp\hat n_\alpha}{\pp\omega} - \frac{\pp \hat n_\alpha}{\pp\sigma} \right) \Bigg/ \left(\frac{\pp^2\hat\Omega}{\pp\sigma^2} - \frac{\left(\frac{\pp^2\hat\Omega}{\pp\sigma\pp\omega}\right)^2}{\frac{\pp^2\hat\Omega}{\pp\omega^2}}\right)\rm,\\
    \frac{\pp\omega}{\pp\hat\mu_\alpha} = &-\left(\frac{\pp \hat n_\alpha}{\pp\omega} + \frac{\pp^2\hat\Omega}{\pp\sigma\pp\omega}\frac{\pp\sigma}{\pp\hat \mu_\alpha}\right) \Bigg/ \frac{\pp^2\hat\Omega}{\pp\omega^2} \rm.
\end{split}
\end{equation}
We note that corresponding derivatives of the mean fields w.r.t. $\hat\mu_\beta$ can be found similarly upon replacing $\alpha \rightarrow \beta$. The above derivatives can be plugged into Eq.~\eqref{eq:x2_ab2}. Now, calculating Eq.~\eqref{eq:x2_ab2} amounts to providing the values of the mean fields and evaluating them numerically. Detailed evaluation of $\pp\sigma/\hat\mu_\alpha$ and $\pp\omega/\hat\mu_\alpha$ is presented in Appendix~\ref{sec:appendix_1}.

\begin{figure*}
    \centering
    \includegraphics[width=.475\linewidth]{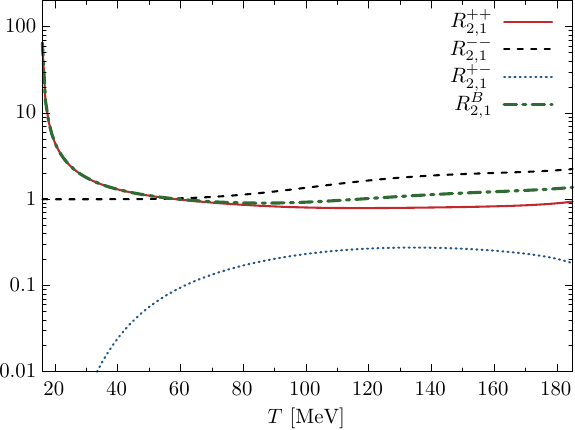}\;\;\;\;
    \includegraphics[width=.475\linewidth]{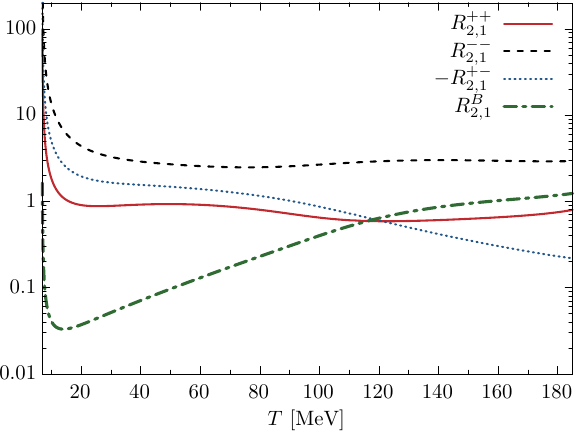}
    \caption{Ratios $R_{2,1}^{\alpha\beta}$ along the crossover liquid-gas (left panel) and chiral (right panel) transition lines defined as the minima of $\pp\sigma/\pp\mu_\pm$ (see text for details). Note that in the right panel $R_{2,1}^{+-}$ is shown with a negative sign.}
    \label{fig:R21_lines}
\end{figure*}

\section{Results}\label{sec:results}

Using Eq.~\eqref{eq:x2_ab2}, we evaluate the susceptibilities of the net number densities for the positive- and negative-parity chiral partners, as well as the correlations among them within the parity doublet model. The results for vanishing baryon chemical potential are shown in Fig.~\ref{fig:x2_T}. The net-baryon susceptibility obtained in the tHRG gas model increases monotonically and does not resemble any critical behavior. This is expected because the partition function of the tHRG gas model is just a sum of ideal, uncorrelated particles [cf. Eq.~\eqref{eq:hrg_thermo}] with vacuum hadron masses. The net-baryon susceptibility obtained in the parity doublet model deviates from the tHRG gas result. The increase around $T_c$ and saturation above it is a bulk consequence of the interplay between critical chiral dynamics with in-medium hadron masses and repulsive interactions~\cite{Marczenko:2021icv}. Around $T_c$, the susceptibilities $\chi_2^{++}$ and $\chi_2^{--}$ develop a swift increase due to chiral symmetry restoration, and therefore the change of their effective masses. They continue to grow at higher temperatures. Up to $T_c$, the correlation $\chi_2^{+-}$ is almost negligible. The reason is that the $N_-$ resonance is thermally suppressed at low temperatures due to its high mass. The correlation only becomes relevant in the vicinity of the chiral crossover, where the negative-parity state becomes swiftly populated. The full net-baryon number susceptibility saturates and gradually decreases to zero at high temperatures due to the non-vanishing correlation between the baryonic chiral partners. We note that $\chi_2^{+-}$ is negative at vanishing $\mu_B$.

Next, we turn to finite baryon chemical potential. In Fig.~\ref{fig:x2_fixed_t_panel}, we show the susceptibilities $\chi_2^{\alpha\beta}$ for different temperatures. At $T=30~$MeV, the net-baryon number susceptibility develops a peak at $\mu_B < 1~$GeV, which is a remnant of the liquid-gas phase transition. At higher chemical potentials, it develops a plateau with a small peak around $\mu_B=1.4~$GeV, which is a remnant of the chiral phase transition. The net-nucleon susceptibility, $\chi_2^{++}$, overlaps with $\chi_2^B$ at small $\mu_B$, which is expected due to thermal suppression of the negative-parity state. On the other hand both $\chi_2^{++}$ and $\chi_2^{--}$ develop strong peaks around $\mu_B\sim 1.4~$GeV. Interestingly, the correlator becomes negative, and  $\chi_2^{+-}$ features a minimum, which is of similar magnitude as the peaks in $\chi_2^{++}$ and $\chi_2^{--}$. Therefore, the negative correlation between the baryonic chiral partners causes the suppression of the net-baryon susceptibility around the chiral crossover [cf.~Eq.~\eqref{eq:x2_sum}]. The structure is similar at $T=50~$MeV.

At low temperature, the liquid-gas and chiral phase transitions are well separated. Higher temperature gives rise to a more complicated structure; the two crossover lines become closer and finally merge (see Fig.~\ref{fig:phase_diag}). This is seen in the bottom panels of Fig.~\ref{fig:x2_fixed_t_panel}. The $\chi_2^B$ features a peak around the chemical potential where the transitions happen. This is not reflected in the individual parity fluctuations; $\chi_2^{--}$ swiftly increase at the chiral crossover, while the correlator $\chi_2^{+-}$ starts to decrease.

In Fig.~\eqref{fig:R21_fixed_t_panel}, we plot the ratios $R_{2,1}^{\alpha \beta}$ for different temperatures. At low temperatures, the ratio $R_{2,1}^{++}$ is sensitive to both liquid-gas and chiral crossovers, while $R_{2,1}^{--}$ is sensitive only to the latter transition. Notably, at the chiral crossover, the peak $R_{2,1}^{--}$ is much stronger than in $R_{2,1}^{++}$. On the other hand, similarly to $\chi_2^B$, the ratio $R_{2,1}^B$ is sensitive to the liquid-gas phase transition, however, it becomes suppressed as compared to $R_{2,1}^{++}$ and $R_{2,1}^{--}$, and the enhancement due to criticality is essentially invisible at the chiral phase boundary. We note that in the close vicinity of the chiral critical endpoint, the $R_{2,1}^B$ ratio indeed shows critical behavior. However, this happens at much lower temperatures. At small $\mu_B$, the ratio $R_{2,1}^{+-}$ is negligibly close to zero and deviates from it only when the negative-parity chiral partner becomes populated, i.e., $R_{2,1}^{--}$ deviates from unity. Its minimum value is obtained in the vicinity of the chiral crossover. This signals the sensitivity of the correlation between the baryonic chiral partners to the onset of chiral symmetry restoration. Interestingly, $R_{2,1}^{--}$ features a well-pronounced peak at high temperatures in the vicinity of the chiral transition, while other quantities do not.

To quantify the differences of fluctuations in the vicinity of the liquid-gas and chiral phase transitions, we calculate the fluctuations as functions of temperature along the trajectories obtained by tracing the remnants of these
two transitions, i.e., the corresponding minima of $\partial \sigma / \partial \mu_+$ and $\partial \sigma / \partial \mu_-$ (see the phase diagram in Fig.~\ref{fig:phase_diag}). The temperature dependence of $R_{2,1}^{\alpha\beta}$ along the remnant of the liquid-gas phase transition is shown in the left panel of Fig.~\ref{fig:R21_lines}. The ratio $R_{2,1}^{++}$ increases toward the critical point of the liquid-gas phase transition, located at  $T\simeq 16~$MeV. On the other hand, $R_{2,1}^{--}$ stays close to unity, due to thermal suppression of the negative-parity nucleon. As a result the  $R_{2,1}^{+-}$ vanishes. Therefore, as the critical point of the liquid-gas phase transition is approached, the system is dominated by the positive-parity state and the fluctuations are entirely due to its contribution. In the right panel of Fig.~\ref{fig:R21_lines}, we show the same quantities along the chiral crossover line. All quantities diverge at the chiral critical point, which is located at $T\simeq 7~$MeV. In this case, the contribution from the negative-parity state is not negligible close to the critical point. Their appearance increases the strength of the correlation between the chiral partners, which becomes large and negatively divergent. In turn, the ratio $R_{2,1}^B$ decreases and starts diverging only in the close vicinity of the chiral critical point. Our results indicate that the net-proton fluctuations do not necessarily reflect the net-baryon fluctuations at the chiral phase boundary.

As we have observed, the susceptibility of the negative-parity state becomes dominant in the vicinity of the chiral critical region. This is even more readily seen in the ratio of the second to first-order susceptibility. Our finding suggests the fluctuations of the negative-parity state provide a good signal to identify the chiral critical point. We remark, however, on the simplified nature of this model calculations. In the current model, the negative-parity state, $N_-(1535)$, is treated as a stable particle rather than resonance with a finite width. However, including finite-width effects in a self-consistent way within the relativistic mean-field approach is not a well-laid procedure. To stipulate more precise expectations of the role of decays in fluctuation observable would require, e.g., to account for the imaginary part of the self-energy of $N_-(1535)$, as done, e.g.,  in~\cite{Suenaga:2017wbb} in the context of dense nuclear matter. This is one of the issues of our forthcoming studies.

\begin{figure}
    \centering
    \includegraphics[width=0.95\linewidth]{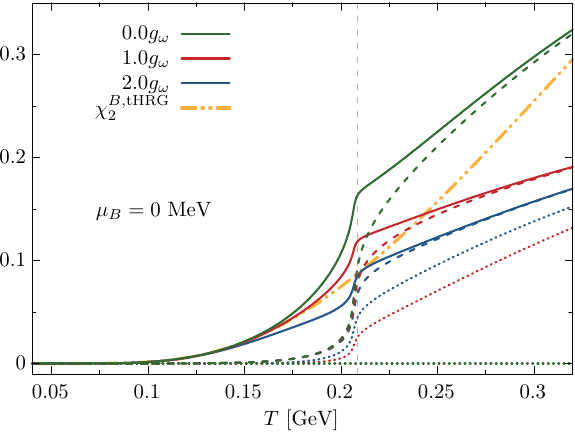}
    \caption{Susceptibilities, $\chi_2^{\alpha\beta}$ at vanishing baryon chemical potential for different values of the repulsive coupling $g_\omega$. The solid, dashed and dotted lines show $\chi_2^{++}$, $\chi_2^{--}$, and $-\chi_2^{+-}$, respectively.}
    \label{fig:x2_T_gw}
\end{figure}

\begin{figure}
    \centering
    \includegraphics[width=0.95\linewidth]{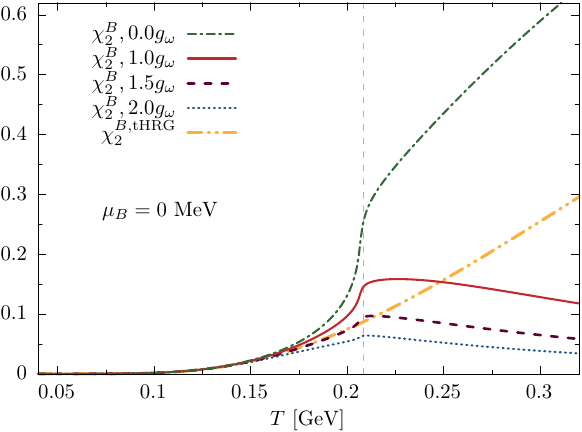}
    \caption{Susceptibility of the net-baryon number density at $\mu_B=0$ as a function of temperature for different values of the repulsive coupling constant $g_\omega$.}
    \label{fig:xB_T_rep}
\end{figure}

\begin{figure}
    \centering
    \includegraphics[width=0.97\linewidth]{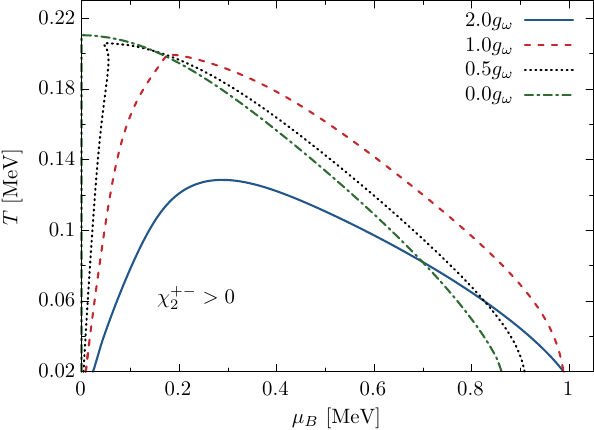}
    \caption{Normalized phase diagram for different values of repulsive coupling $g_\omega$. The lines correspond to vanishing correlator $\chi_2^{+-}$ and the areas enclosed by them show regions where $\chi_2^{+-}>0$.}
    \label{fig:diagram_corr}
\end{figure}

\section{Effect of repulsion}

The repulsive interactions have little to none effect on the chiral crossover transition at small baryon chemical potentials. This is expected due to the vanishing of the $\omega$ mean field at $\mu_B=0$. In Fig.~\ref{fig:x2_T_gw}, we show the susceptibilities for different values of the repulsive coupling $g_\omega$ and other parameters kept fixed at $\mu_B=0$. As expected, for vanishing coupling, fluctuations are the largest, and the correlator $\chi_2^{+-}$ vanishes. As the value of $g_\omega$ increases, the fluctuations of the positive- and negative-parity state become suppressed. At the same time, finite $g_\omega$ implies finite correlations, which otherwise vanish at $\mu_B=0$. With increasing the coupling, the correlations become more negative, further suppressing the total net-baryon number fluctuations. Thus, it is the correlation between the baryonic chiral partners that non-trivially modifies the net-baryon number fluctuations.

While in-medium effects due to chiral symmetry restoration may spoil agreement between the HRG model and LQCD results on the second-order susceptibilities, it can be potentially restored by tuning the strength of repulsive interactions. This can be deduced from Fig.~\ref{fig:xB_T_rep}, where we compare susceptibilities of the net-baryon number density $\chi_2^B$ for different values of the repulsive coupling constant. For vanishing repulsive coupling, the susceptibility swiftly increases and overestimates the tHRG gas result in the vicinity of the chiral crossover. In general, as the repulsive coupling increases, the fluctuations tend to decrease~\cite{Kunihiro:1991qu}. For twice the value of the original coupling, the susceptibility already underestimates the tHRG gas result. Therefore, by choosing value somewhere in between, the in-medium effects would cancel out and the agreement with tHRG gas fluctuations would be restored.

To see the effect of the repulsion on the phase structure, in Fig.~\ref{fig:diagram_corr}, we plot the phase diagram of the model in the $T-\mu_B$ plane for different values of the repulsive coupling $g_\omega$. In general, smaller repulsive coupling yields the region where $\chi_2^{+-}>0$ more tilted to the left. Nevertheless, the qualitative structure remains the same, regardless of the presence of the repulsive forces. We note that in Fig.~\ref{fig:diagram_corr}, we do not show results for $T<20$, where the liquid-gas and chiral transitions become of first-order and additional effects, such as non-equilibrium spinodal decomposition have to be addressed. These interesting effects have been already explored in the context of the Nambu--Jona-Lasinio model~\cite{Sasaki:2007db, Sasaki:2007qh}. This is, however, beyond the scope of the current work and we plan to elaborate on this elsewhere.

Now, we focus on the properties of the correlator, in particular on the change of its sign at finite chemical potential. Because the qualitative behavior of the correlator does not depend on the repulsive interactions, we consider $g_\omega=0$ and neglect the vector channel. Then, the correlator in Eq.~\eqref{eq:chi_2_ab} simplifies to the following
\begin{equation}
  \chi_2^{\alpha\beta} = \frac{1}{\frac{\partial^2 \hat \Omega}{ \partial \sigma^2 } } \frac{\partial \hat n_\alpha}{\partial \sigma} \frac{\partial \hat n_\beta}{\partial \sigma} = \frac{1}{\frac{\partial^2 \hat\Omega}{\partial \sigma^2}} \frac{\partial \hat n_\alpha}{\partial m_\alpha} \frac{\partial \hat n_\beta}{\partial m_\beta} \frac{\pp m_\alpha}{\pp \sigma}\frac{\pp m_\beta}{\pp \sigma} \rm. 
\end{equation}
Since the curvature, $\frac{\partial^2 \hat \Omega}{ \partial \sigma^2 }>0$ is positive, the sign change in the correlator at finite baryon chemical potential is related to the change of the sign of $\pp m_\pm/ \pp \sigma$. From Eq.~\eqref{eq:m_ds}, one sees that at $\sigma_{\rm min}$, the correlator $\chi_2^{+-}$ changes sign, while $\chi_2^{++}$ and $\chi_2^{--}$ stay positive. Indeed, we have confirmed this numerically for vanishing repulsive interactions. Nevertheless, in a more realistic scenario with repulsive interactions, they provide additional sources of negative correlations. This is seen in Fig.~\ref{fig:diagram_corr}, where the vanishing $\chi_2^{+-}$ lines lie at $\mu_B<1~$GeV, where $\sigma > \sigma_{\rm min}$ (compare with Fig.~\ref{fig:phase_diag}). Therefore, the overall behavior of the correlator is given by a non-trivial interplay between chiral symmetry restoration and repulsive interactions.

\section{Conclusions}\label{sec:conclusions}

We have investigated the net-baryon number density fluctuations and discussed the qualitative role of chiral criticality of hadronic matter at finite temperature and baryon chemical potential. In particular, we have studied for the first time the susceptibilities of the positive- and negative-parity chiral partners, as well as their correlations. To this end, we have used the parity doublet model in the mean-field approximation. We have analyzed the thermodynamic properties and the susceptibility of the net-baryon number. 

We have confirmed that in the vicinity of the liquid-gas phase transition, the net baryon number density is dominated by the contribution of the positive-parity state. In contrast, this does not need to be the case at the boundary of the chiral crossover. We find that there,   the fluctuations of the net-baryon number density are suppressed, compared to the positive-parity state fluctuations (i.e. net-nucleon). This qualitative difference is not only due to the presence of the negative-parity state but largely due to the non-trivial correlation between the chiral partners.

The qualitative differences in the net-nucleon and net-baryon fluctuations can also be useful in searching for possible critical points in the QCD phase diagram. In particular, our results bring significant and nontrivial differences in the critical behavior of the net-nucleon fluctuations in the vicinity of the liquid-gas and chiral phase transitions. This strongly suggests that in order to fully interpret the critical properties of the matter created in heavy-ion collisions, especially in the forthcoming large-scale nuclear experiments FAIR at GSI and NICA in Dubna, it is essential to consistently incorporate and understand the chiral in-medium effects carried by the baryonic parity partners and their correlations. 

To reach further theoretical insights and understanding of the QCD phase diagram, it is important to determine correlations between baryonic chiral partners of opposite parity in lattice QCD calculations. Furthermore, to elaborate on the relationship between net-nucleon and net-baryon fluctuations, it is desirable to perform more refined calculations of the higher-order susceptibilities and their ratios. It is also useful to understand the role of finite width and decay properties of the negative parity states on the fluctuation observables. Work in these directions is in progress and will be reported elsewhere.

\section*{Acknowledgements}
M.M. and K.R. acknowledge fruitful discussions and helpful suggestions from Bengt Friman and Nu Xu. This work is supported partly by the Polish National Science Centre (NCN) under OPUS Grant No. 2022/45/B/ST2/01527 (K.R. and C.S.), Preludium Grant No. 2017/27/N/ST2/01973 (M.M.), and the program Excellence Initiative–Research University of the University of Wroc\l{}aw of the Ministry of Education and Science (M.M.). The work of C.S. was supported in part by the World Premier International Research Center Initiative (WPI) through MEXT, Japan. K.R. also acknowledges the support of the Polish Ministry of Science and Higher Education. V.K. acknowledges the support of the University of Wroc\l{}aw within the IDUB visiting professor program. V.K also would like to thank GSI and the Institute for Nuclear Theory at the University of Washington for their kind hospitality and stimulating research environment. V.K. has been supported by the U.S. Department of Energy, Office of Science, Office of Nuclear Physics, under contract number DE-AC02-05CH11231, by the INT's U.S. Department of Energy grant No. DE-FG02-00ER41132, and by the ExtreMe Matter Institute EMMI at the GSI Helmholtzzentrum für Schwerionenforschung, Darmstadt, Germany. M.M. acknowledges the support of the European Union’s Horizon 2020 research and innovation program under grant agreement No 824093 (STRONG-2020).

\appendix

\section{Parity doublet model Lagrangian}
\label{sec:PDM_appendix}

To explore the criticality linked to the chiral symmetry restoration and its consequences in baryonic matter at finite temperature and density we adopt the parity doublet model~\cite{Detar:1988kn, Jido:1999hd, Jido:2001nt}. In the conventional Gell-Mann--Levy model of mesons and nucleons~\cite{GellMann:1960np}, the nucleon mass is entirely generated by the non-vanishing expectation value of the sigma field. Thus, the nucleon inevitably becomes massless when the chiral symmetry is restored. This is led by the particular chirality assignment to the nucleon parity doublers, where the nucleons are assumed to be transformed in the same way as the quarks are under chiral rotations.

More general allocation of the left- and right-handed chiralities to the nucleons, the mirror assignment, was proposed in~\cite{Detar:1988kn}. This allows an explicit mass term for the nucleons, and consequently, the nucleons stay massive at the chiral restoration point. For more details, see Refs.~\cite{Detar:1988kn,Jido:1999hd,Jido:2001nt}.

In the mirror assignment, under \mbox{$SU(2)_L \times SU(2)_R$} rotation, two chiral fields $\psi_1$ and $\psi_2$ are transformed as follows:
\begin{equation}\label{eq:mirror_assignment}
\begin{split}
    \psi_{1L} \rightarrow L\psi_{1L}, \;\;\;\; \psi_{1R} \rightarrow R\psi_{1R}\textrm, \\
    \psi_{2L} \rightarrow R\psi_{2L}, \;\;\;\; \psi_{2R} \rightarrow L\psi_{2R}\textrm,
\end{split}
\end{equation}
where $\psi_i = \psi_{iL} + \psi_{iR}$, $L \in SU(2)_L$ and $R \in SU(2)_R$. In this work, we consider a system with $N_f = 2$, hence, relevant for this study are the lowest nucleons and their chiral partners. The hadronic degrees of freedom are coupled to the chiral fields $(\sigma,~\pi)$, and the iso-singlet vector field $\omega_\mu$. The nucleon part of the Lagrangian in the mirror model reads
\begin{equation}\label{eq:doublet_lagrangian}
\begin{split}
    \mathcal{L}_N &= i\bar\psi_1\slashchar\partial\psi_1 + i\bar\psi_2\slashchar\partial\psi_2 + m_0\left(  \bar\psi_1\gamma_5\psi_2 - \bar\psi_2\gamma_5\psi_1 \right) \\
    &+ g_1\bar\psi_1 \left( \sigma + i\gamma_5 \boldsymbol\tau \cdot \boldsymbol\pi \right)\psi_1 + g_2\bar\psi_2 \left( \sigma - i\gamma_5 \boldsymbol\tau \cdot \boldsymbol\pi \right)\psi_2 \\
    &-g_\omega\bar\psi_1\slashchar\omega\psi_1 - g_\omega\bar\psi_2\slashchar\omega\psi_2 \textrm,
\end{split}
\end{equation}
where $g_1$, $g_2$, and $g_\omega$ are the baryon-to-meson coupling constants and $m_0$ is a mass parameter. Note that we assume the same vector coupling strength for both parity partners.

The mesonic part of the Lagrangian reads
\begin{equation}
\begin{split}
    \mathcal{L}_M = \frac{1}{2} \left( \partial_\mu \sigma\right)^2 + \frac{1}{2} \left(\partial_\mu \boldsymbol\pi \right)^2 - \frac{1}{4} \left( \omega_{\mu\nu}\right)^2-V_\sigma - V_\omega \textrm,
\end{split}
\end{equation}
where $\omega_{\mu\nu} = \partial_\mu\omega_\nu - \partial_\nu\omega_\mu$ is the field-strength tensor of the vector field, and the potentials are defined in Eq.~\eqref{eq:potentials_parity_doublet}.

The full Lagrangian of the parity doublet model is given by
\begin{equation}
    \mathcal L = \mathcal L_N + \mathcal L_M\textrm.
\end{equation}

The mass eigenstates of the parity partners, $N_\pm$ are obtained by diagonalizing the mass matrix for $\psi_1$ and $\psi_2$:
\begin{equation}
    \begin{pmatrix} N_+ \\ N_- \end{pmatrix} = \frac{1}{\sqrt{2\cosh \delta}} \begin{pmatrix}
        e^{\delta/2} & \gamma_5 e^{-\delta/2}\\
        \gamma_5 e^{-\delta/2} & -e^{\delta/2}
    \end{pmatrix} 
    \begin{pmatrix}
        \psi_1 \\
        \psi_2
    \end{pmatrix} \rm,
\end{equation}
where $\sinh \delta = -a\sigma / 2m_0$, and $a=g_1+g_2$. In the diagonal basis, the masses of the positive- and negative-parity baryonic chiral partners, $N_\pm$, are given by 
\begin{equation}\label{eq:doublet_masses2}
    m_\pm = \frac{1}{2}\left(\sqrt{a^2\sigma^2 + 4m_0^2} \mp b\sigma\right) \textrm,
\end{equation}
where $b=g_1-g_2$. From Eq.~(\ref{eq:doublet_masses2}), it is clear that, in contrast to the naive assignment under chiral symmetry, the chiral symmetry breaking generates only the splitting between the two masses. When the symmetry is restored, the masses become degenerate, $m_\pm(\sigma\rightarrow0) \rightarrow m_0$.

\section{Cumulants of the net-baryon number}
\label{sec:appendix_0}

In the following, we present a derivation of the second-order cumulants of the positive/negative parity baryons and their correlator.

We start by recalling that $\delta N_B = N_B - \langle N_B \rangle$. From this, it follows that the variance can be written as
\begin{equation}\label{eq:variance_app}
\begin{split}
    \langle \delta N_B \delta N_B\rangle &= \langle \delta N_B^2\rangle = \langle \left(N_B - \langle N_B\rangle \right)^2 \rangle \\
    &= \langle N_B^2\rangle + \langle \langle N_B \rangle^2\rangle - 2\langle N_B\langle N_B\rangle\rangle \\
    &= \langle N_B^2\rangle + \langle N_B \rangle^2 - 2\langle N_B\rangle^2 \\
    &= \langle N_B^2\rangle - \langle N_B\rangle^2 \rm,\\
\end{split}
\end{equation}
which is a standard form of the variance of the distribution of variable $N_B$. 

For a system consisting of $N_B = N_+ + N_-$ baryons, Eq.~\eqref{eq:variance_app} can be rewritten as
\begin{equation}
\begin{split}
    \langle \delta N_B \delta N_B\rangle =&\;\langle (N_+ + N_-)^2\rangle - \langle N_+ + N_-\rangle^2 \\
    =&\; \langle N_+^2\rangle + \langle N_-^2\rangle + 2\langle N_+N_-\rangle \\
    &- \langle N_+\rangle^2 - \langle N_-\rangle^2 - 2\langle N_+\rangle \langle N_-\rangle \\
    \equiv &\; \kappa_2^{++} + \kappa_2^{--} + 2 \kappa_2^{+-}\rm ,
\end{split}
\end{equation}
where
\begin{equation}
    \kappa_2^{\alpha\beta} = \langle N_\alpha N_\beta \rangle - \langle N_\alpha \rangle \langle N_\beta \rangle = \langle \delta N_\alpha\delta N_\beta \rangle \rm,
\end{equation}
for $\alpha,\beta = \pm$ referring to baryons with positive/negative parity. $\kappa_2^{++}$, $\kappa_2^{--}$ are the cumulants in the individual parity channels and $\kappa_2^{+-}$ is their correlator.

In the grand canonical ensemble the partition function $\mathcal Z=\mathcal Z\left(T,V,\mu_B\right)$ is given as
\begin{equation}\label{eq:part_func_app}
    \mathcal Z = \sum_{N_B} e^{\mu_B N_B/T} \mathcal Z_{\rm C} = \sum_{N_B} e^{\hat \mu_B N_B} \mathcal Z_{\rm C}\rm,
\end{equation}
where $\mathcal Z_{\rm C} = \mathcal Z_{\rm C}(T,V,N_B)$ is the canonical partition function and $\hat \mu_x = \mu_x /T$. 

The cumulants are defined as derivatives of the partition function w.r.t. to a chemical potential. For a system composed of $N_B = N_+ + N_-$, we may rewrite the exponent as $(\hat\mu_B N_+ + \hat\mu_B N_-)$. Finally, one may relabel the chemical potentials as $(\hat\mu_+ N_+ + \hat\mu_- N_-)$, keeping in mind that $\hat\mu_+ = \hat\mu_- = \hat\mu_B$. This trick allows us to take derivatives directly w.r.t to individual chemical potentials, $\hat\mu_\pm$. The first derivative w.r.t. to chemical potential, $\hat\mu_\alpha$,
\begin{equation}
    \frac{\dd \log \mathcal Z}{\dd \hat\mu_\alpha}\Bigg|_T = \frac{1}{\mathcal Z}\sum_{N_+,N_-} N_\alpha e^{\hat\mu_B N_B} \mathcal Z_{\rm C} = \langle N_\alpha \rangle \equiv \kappa_1^{\alpha} \rm ,
\end{equation}
gives the first-order cumulant, i.e., the net number of baryons of $N_\pm$. Consequently, taking the second derivative w.r.t. $\hat\mu_\beta$, 
\begin{equation}
\begin{split}
    &\frac{\dd^2 \log \mathcal Z}{\dd \hat\mu_\alpha \dd \hat\mu_\beta}\Bigg|_T = \\
    =&-\frac{1}{\mathcal Z^2}\sum_{N_+,N_-} N_\alpha e^{\hat\mu_B N_B} \mathcal Z_{\rm C} \sum_{N_+,N_-} N_\beta e^{\hat\mu_B N_B} \mathcal Z_{\rm C}\\
    &+ \frac{1}{\mathcal Z} \sum_{N_+, N_-} N_\alpha N_\beta e^{\hat\mu_B N_B}\mathcal Z_{\rm C} \\
    =& \langle N_\alpha N_\beta\rangle - \langle N_\alpha \rangle \langle N_\beta \rangle \equiv \kappa_2^{\alpha\beta}\rm ,
\end{split}
\end{equation}
which gives the second-order cumulants of $N_\pm$ and their correlator.

\section{Detailed evaluation of the susceptibilities \texorpdfstring{$\chi_2^{\alpha\beta}$}{}}
\label{sec:appendix_1}

In this appendix, we show in detail the evaluation of the susceptibilities $\chi_2^{\alpha\beta}$ given in Eq.~\eqref{eq:chi_2_ab},
\begin{equation}\label{eq:chi_2_ab_appendix}
    \chi_2^{\alpha\beta} = -\frac{\dd^2 \hat\Omega}{\dd \hat\mu_\alpha \dd\hat\mu_\beta}\Bigg|_{T} \rm,
\end{equation}
where $\hat \Omega = \Omega /T^4$ and $\hat \mu_x = \mu_x / T$. We start by taking the derivatives explicitly and remembering that $\hat \Omega$ is a function of the mean fields. One gets the following
\begin{equation}\label{eq:x2_ab2_appendix}
\begin{split}
    \chi_2^{\alpha\beta} = &-\frac{\pp \sigma}{\pp \hat\mu_\beta} \left( \frac{\pp^2 \hat\Omega}{\pp \sigma^2}\frac{\pp \sigma}{\pp \hat\mu_\alpha} + \frac{\pp^2\hat\Omega}{\pp\sigma\pp\omega}\frac{\pp\omega}{\pp\hat\mu_\alpha} + \frac{\pp^2 \hat \Omega}{\pp\hat\mu_\alpha\pp\sigma} \right)\\
    &-\frac{\pp\omega}{\pp\hat\mu_\beta} \left( \frac{\pp^2\hat\Omega}{\pp\omega^2}\frac{\pp \omega}{\pp\hat\mu_\alpha} + \frac{\pp^2\Omega}{\pp\sigma\pp\omega}\frac{\pp\sigma}{\pp\hat\mu_\alpha} + \frac{\pp^2 \hat \Omega}{\pp\hat\mu_\alpha\pp\omega}\right)\\
    &-\frac{\pp\sigma}{\pp\hat \mu_\alpha}\frac{\pp^2 \hat \Omega}{\pp\hat\mu_\beta\pp\sigma} - \frac{\pp\omega}{\pp\hat \mu_\alpha}\frac{\pp^2\hat \Omega}{\pp\hat\mu_\beta\pp\omega} - \frac{\pp^2 \hat \Omega}{\pp\hat\mu_\alpha\pp\hat \mu_\beta} \rm.
    \end{split}
\end{equation}
To evaluate this expression, we need to calculate the derivatives of the mean fields, $\pp\sigma/\hat\mu_\alpha$ and $\pp\omega/\hat\mu_\alpha$\footnote{We note that corresponding derivatives of the mean fields w.r.t. $\hat\mu_\beta$ can be found similarly upon replacing $\alpha \rightarrow \beta$.}. This can be done by taking advantage of the stationary conditions and differentiating the gap equations:
\begin{equation}\label{eq:gaps_appendix}
\begin{split}
    &\frac{\dd}{\dd \hat\mu_\alpha}\left(\frac{\pp \hat\Omega}{\pp \sigma}\right)\Bigg|_{T} = \frac{\pp^2 \hat \Omega}{\pp\sigma^2}\frac{\pp \sigma}{\pp\hat \mu_\alpha} + \frac{\pp^2 \hat \Omega}{\pp\omega\pp\sigma}\frac{\pp \omega}{\pp\hat \mu_\alpha} + \frac{\pp^2 \hat \Omega}{\pp\hat\mu_\alpha \pp\sigma} = 0\rm,\\
    &\frac{\dd}{\dd \hat\mu_\alpha}\left(\frac{\pp \hat\Omega}{\pp \omega}\right)\Bigg|_{T} = \frac{\pp^2 \hat \Omega}{\pp\sigma\pp\omega}\frac{\pp \sigma}{\pp\hat \mu_\alpha} + \frac{\pp^2 \hat \Omega}{\pp\omega^2}\frac{\pp \omega}{\pp\hat \mu_\alpha} + \frac{\pp^2 \hat \Omega}{\pp\hat\mu_\alpha \pp\omega} = 0 \rm.
\end{split}
\end{equation}
Before proceeding, we introduce the following shorthand notation:
\begin{equation}
    \hat \Omega_{\phi\eta} = \frac{\pp^2\hat \Omega}{\pp\phi\pp\eta}\rm,
\end{equation}
and
\begin{equation}
    \hat \Omega_{\alpha\phi} = \frac{\pp^2\hat \Omega}{\pp\hat\mu_\alpha\pp\phi}\rm, \;\;\;\;\phi_\alpha = \frac{\pp \phi}{\pp \hat\mu_\alpha}\rm,
\end{equation}
where $\phi$ and $\eta$ denote the mean fields $\sigma$ and $\omega$, and $\alpha = \pm$ denotes the parity partners.

Applying the simplified notation to Eqs.~\eqref{eq:x2_ab2_appendix} and \eqref{eq:gaps_appendix} gives 
\begin{equation}\label{eq:x2_ab2_appendix2}
\begin{split}
    \chi_2^{\alpha\beta} = &-\sigma_\beta\left( \hat\Omega_{\sigma\sigma}\sigma_\alpha + \hat\Omega_{\sigma\omega}\omega_\alpha + \hat\Omega_{\alpha\sigma} \right)\\
    &-\omega_\beta\left(\hat\Omega_{\omega\omega}\omega_\alpha + \hat\Omega_{\sigma\omega}\sigma_\alpha + \hat\Omega_{\alpha\omega}\right)\\
    &-\sigma_{\alpha}\hat\Omega_{\beta\sigma} - \omega_\alpha\hat\Omega_{\beta\omega} - \hat\Omega_{\alpha\beta} \rm,
    \end{split}
\end{equation}
and 
\begin{align}
    &\hat\Omega_{\sigma\sigma}\sigma_\alpha + \hat\Omega_{\sigma\omega}\omega_\alpha + \hat\Omega_{\alpha\sigma} = 0\rm,\label{eq:gaps_appendi_s}\\
    &\hat\Omega_{\sigma\omega}\sigma_\alpha + \hat\Omega_{\omega\omega}\omega_\alpha + \hat\Omega_{\alpha\omega}= 0 \rm,\label{eq:gaps_appendi_w}
\end{align}
respectively.

Isolating $\omega_\alpha$ in Eq.~\eqref{eq:gaps_appendi_w} gives
\begin{equation}\label{eq:w_a_appendix}
    \omega_\alpha = - \left(\hat\Omega_{\alpha\omega} + \hat\Omega_{\sigma\omega}\sigma_\alpha\right) \Big/ \hat\Omega_{\omega\omega}\rm.
\end{equation}
Next, substituting Eq.~\eqref{eq:w_a_appendix} into Eq.~\eqref{eq:gaps_appendi_s}, we get
\begin{equation}
    \hat\Omega_{\sigma\sigma}\sigma_\alpha + \hat\Omega_{\alpha\sigma} - \frac{\hat\Omega_{\sigma\omega}}{\hat\Omega_{\omega\omega}}\left(\hat\Omega_{\alpha\omega} + \hat\Omega_{\sigma\omega}\sigma_\alpha\right)  = 0\rm.
\end{equation}
Isolating $\sigma_\alpha$ in the above expression yields
\begin{equation}\label{eq:s_a_appendix2}
\sigma_\alpha = \left(\hat\Omega_{\alpha\omega}\frac{\hat\Omega_{\sigma\omega}}{\hat\Omega_{\omega\omega}} -\hat\Omega_{\alpha\sigma} \right) \left(\hat\Omega_{\sigma\sigma} - \frac{\hat\Omega_{\sigma\omega}^2}{\hat\Omega_{\omega\omega}}\right)^{-1} \rm,
\end{equation}
which can be plugged into Eq.~\eqref{eq:w_a_appendix} to get
\begin{equation}\label{eq:w_a_appendix2}
    \omega_\alpha = \left[\hat\Omega_{\sigma\omega}\frac{\hat\Omega_{\alpha\sigma}\hat\Omega_{\omega\omega}-\hat\Omega_{\alpha\omega}\hat\Omega_{\sigma\omega}  }{\hat\Omega_{\sigma\sigma}\hat\Omega_{\omega\omega}- \hat\Omega_{\sigma\omega}^2}-\hat\Omega_{\alpha\omega} \right] \Big/ \hat\Omega_{\omega\omega} \rm.
\end{equation}
Once the gap equations are solved for $\sigma$ and $\omega$ at given $T$ and $\mu_B$, Eqs.~\eqref{eq:s_a_appendix2} and \eqref{eq:w_a_appendix2} can be evaluated numerically.

To further simplify the above expressions, let us recall the definition of the net density,
\begin{equation}
    \hat n_\alpha = - \frac{\pp \hat \Omega}{\pp \hat\mu_\alpha} = -\hat\Omega_{\alpha} \rm.
\end{equation}
Which allows us to write the mixed derivative of the thermodynamic potential 
\begin{equation}
    \hat\Omega_{\alpha\phi} = - \hat n_{\alpha\phi} \rm.
\end{equation}
Now, Eqs.~\eqref{eq:s_a_appendix2} and \eqref{eq:w_a_appendix2} can be written as
\begin{align}
\sigma_\alpha &= \frac{\hat n_{\alpha\sigma}\hat\Omega_{\omega\omega} -\hat n_{\alpha\omega}\hat\Omega_{\sigma\omega} }{\hat\Omega_{\sigma\sigma}\hat\Omega_{\omega\omega} - \hat\Omega_{\sigma\omega}^2} \rm, \label{eq:s_a_appendix3}\\
\omega_\alpha &= \frac{\hat n_{\alpha\omega}}{\hat\Omega_{\omega\omega}} + \frac{\hat\Omega_{\sigma\omega}} {\hat\Omega_{\omega\omega}} \frac{\hat n_{\alpha\sigma}\hat\Omega_{\omega\omega} -\hat n_{\alpha\omega}\hat\Omega_{\sigma\omega} }{\hat\Omega_{\sigma\sigma}\hat\Omega_{\omega\omega} - \hat\Omega_{\sigma\omega}^2}\rm.\label{eq:w_a_appendix3}
\end{align}

Finally, the expression for the susceptibilities can be rewritten as follows
\begin{equation}\label{eq:x2_ab2_appendix3}
\begin{split}
    \chi_2^{\alpha\beta} = &-\sigma_\beta\left(\hat\Omega_{\sigma\sigma}\sigma_\alpha + \hat\Omega_{\sigma\omega}\omega_\alpha - \hat n_{\alpha\sigma} \right)\\
    &-\omega_\beta\left(\hat\Omega_{\omega\omega}\omega_\alpha + \hat\Omega_{\sigma\omega}\sigma_\alpha - \hat n_{\alpha\omega}\right)\\
    &+\sigma_{\alpha}\hat n_{\beta\sigma} + \omega_\alpha\hat n_{\beta\omega} + \hat n_{\alpha\beta} \rm.
    \end{split}
\end{equation}
Eqs.~\eqref{eq:s_a_appendix3} and \eqref{eq:w_a_appendix3} can be inserted into Eq.~\eqref{eq:x2_ab2_appendix3} to obtain the susceptibilities in the individual parity channels and the correlator. We note that at the end of the analytical evaluation, the chemical potentials $\hat\mu_\alpha = \hat\mu_\beta = \hat\mu_N$.

\bibliography{biblio}

\end{document}